\documentclass[pre,twocolumn,showpacs,showkeys,preprintnumbers,amsmath,amssymb,floatfix]{revtex4}

\usepackage{graphicx}
\usepackage{dcolumn}
\usepackage{bm}
\usepackage{amsmath}
\usepackage{amsfonts}
\usepackage{amssymb}
\usepackage{amsthm}
\usepackage{newlfont}
\usepackage{graphicx}
\usepackage{amscd}

\begin{document}


\title{
Analytic approach to stochastic cellular automata:\\
exponential and inverse power distributions out of \\
Random Domino Automaton
}

\author{Mariusz Bia{\l}ecki}
\email{bialecki@igf.edu.pl}
 
\author{Zbigniew Czechowski}%
 \email{zczech@igf.edu.pl}
\affiliation{%
Institute of Geophysics, Polish Academy of Sciences \\
ul.~Ks. Janusza 64, 01-452 Warszawa, Poland}%

\date{\today}

\begin{abstract}
We introduce the stochastic domino cellular automaton model exhibiting avalanches. 
Depending of the choice of the parameters, the model covers wide range of properties:
various types of exponential and long tail (up to inverse-power) distributions of avalanches are observed.
The stationary state of automaton is described by a set of nonlinear discrete equations
derived in an exact way from elementary combinatorial arguments. 
These equations allow to derive formulas explaining both various exponential and inverse power distributions
relating them to values of the parameters.  
The exact relations between the state variable of the model (moments)
are derived in two ways: from direct arguments and from the set of equations. 
Excellent agreement of the obtained analytical results with numerical simulations is observed. 
\end{abstract}


\maketitle


\section{Introduction}

There is a significant interest in constructing simple stochastic models with avalanches reflecting properties 
of various natural phenomena, but only part of them have the advantage of being solvable.
One way to make models analytically tractable is to use the mean-field approximation.
However, requirements of exact results can force further simplifications in a model.
For example, in terms of self-organized criticality, in spite of the fact that the mean field theory of such critical phenomena was already proposed \cite{TangBak-mf}, Dhar made further simplifications for distinguished BTW model \cite{DharExact} 
(a unified view of mean-field picture of stochastic self-organized critical models is presented in \cite{Vespignani}).
We point out also that considering structure of abelian algebras 
leads to analytical results in some sand-pile and related stochastic models \cite{DharSanpile, Alcaraz-DirAbAlg}.
Solvable simple models can also be constructed in the field of directed percolation; for an extremely simple one, see \cite{Gaveau-mfsoc}.

Another way towards solvability is to consider stochastic properties as non-essential and  
study deterministic models; see such an approach for sand-pile in \cite{detSandpile}.
In the context of deterministic cellular automata, an analytic approach related to solvability is 
investigated as part of the theory of integrable systems and may employ sophisticated methods.
\cite{TS-aut, TTMS, TokiLNP, DBK, BD-hyp, BiaRIMS}. We underline that construction of integrable cellular automata is definitely not an easy task.

Our aim here is to follow the ideas of analytic approach and apply them to construction of {\it{stochastic}} automata.
In the article we propose and analyse in a direct elementary way the Random Domino Automaton (RDA) - a new slowly driven systems exhibiting avalanche phenomena. We prefer here an elementary self-contained approach to the description
of the automaton and we make no use of applicable Markov processes terminology. We stress here, all 
results, except of equation \eqref{eq:powerlaw} (see \cite{BiaMotzkin} for details), have elementary derivation in the text below. On the other hand, in spite of its simple formulation and being analytically tractable, 
automaton covers wide range of behaviours depending on the choice of the parameters. 
The application of the RDA model for studying Ito equation is investigated in our parallel papers \cite{CzBiaTL, CzBiaEF}. 

An inspiration for defining the rule for the domino automaton comes from very simplified view of earthquakes.
It corresponds to two tectonic plates moving with relative constant velocity. The wedge may be irregularly rough, and  relative motion can be locked in some places, producing stress accumulation. Beyond some threshold of stress, 
a relaxation took place. The size of relaxation depends on the nearby accumulated stress.  
RDA is inspired by earthquakes; however, a direct reproducing of realistic-like behaviour 
of such complicated phenomena is obviously beyond its scope in the present form. 
Nevertheless, construction and analysis of models is one of primary aims in geosciences \cite{SornetteCritical, LNP705}.
There are many very simplified cellular automata models focusing on specific features of the investigated behaviour in the field. Here we point out a sequence of papers \cite{PachMin, PachTect, Pach08}, where some interesting cellular automata 
models were presented.

Finally, as an unexpected property we mention an interesting link between stochastic cellular automata
and integer sequences (see \cite{SloaneEnc} or The On-Line Encyclopedia of Integer Sequences).
It is known, how to obtain Catalan numbers out of the bond directed percolation on a square lattice \cite{Inui-Catalan}.
Our cellular automaton in specific case leads to Motzkin numbers (for details see \cite{BiaMotzkin}).

Plan of the article is as follows.
In Section \ref{sec:autdef} we introduce a definition of random domino automaton with rebound parameters.
Section \ref{sec:eqs} contains full  combinatorial 
derivation of equations for the distribution of clusters, which describes stationary state of the automaton. 
Several exact formulas - balance relations, average size of clusters and avalanche, general formula for all moments - are displayed. We finish with introducing special form of rebound parameters, which leads to two distinguished
special cases studied in details in next Section \ref{sec:speccases}. These cases correspond to exponential 
and inverse-power distributions as shown below. The obtained analytical results are compared with simulations.
Section \ref{sec:concl} gives a resume of results presented in the article as well as directions for future work.
Appendix \ref{app:empty_cl_derv} contains derivation of complementary set of equations, describing empty clusters 
of the automaton.

\section{Definition of random domino automaton} \label{sec:autdef}
In the random domino automaton model, the space consists of  discrete number $N$ of cells on a line, and we assume periodic boundary conditions. Each cell may be in one of two states: empty state, when it is empty, or occupied state, when contains a ball.
The evolution of the automaton is given by the following update procedure performed in each discrete time step.  
A ball is added to the system to the randomly chosen cell and we assume each cell to be equally possible. If the chosen cell is empty, there are two possibilities: it becomes occupied with probability $\nu$ or the ball is rebounded with probability $(1-\nu)$ leaving the state of the automaton unchanged. 
If the chosen place is already occupied, there are also two possibilities: the ball is rebounded with probability $(1-\mu)$ or with probability $\mu $ the incoming ball triggers a relaxation. 
By relaxation we mean: balls from the chosen cell and from all its adjacent occupied cells are removed. 
Thus, the relaxation produces an avalanche of size equal to the number of cells changing their state. Then the update procedure repeats in the next time step. All possibilities are shown schematically on the diagram below.
\begin{center} 
	\begin{tabular}{c|c|c|c|c|c|c|c|c|c|c|c|c}
	
\multicolumn{6}{c}{ \ }  & \multicolumn{1}{c}{${\bullet} $} &  \multicolumn{6}{c}{ \ } \\
\multicolumn{5}{c}{ \ } & \multicolumn{1}{c}{ \ $\swarrow$ } & \multicolumn{1}{c}{ \ } & 
\multicolumn{1}{c}{ $\searrow $ \ } & \multicolumn{5}{c}{ \  } \\
\multicolumn{6}{r}{ $ (1-\rho)$ }  & \multicolumn{2}{c}{ \ } & \multicolumn{1}{l}{ $\rho$ } & \multicolumn{4}{c}{ \ } \\
\multicolumn{3}{c}{ \ } & \multicolumn{1}{c}{ \ $\swarrow$ } & \multicolumn{5}{c}{ \ } & 
\multicolumn{1}{c}{ $\searrow $ \ } & \multicolumn{3}{c}{ \  } \\

\multicolumn{2}{c}{ \ } & { \ldots } &  $ \ $  & \multicolumn{1}{c}{ \ldots } & \multicolumn{3}{c}{ \ } & 
\ldots &  $ \bullet $  & \multicolumn{1}{c}{ \ldots }  & \multicolumn{2}{c}{ \ }	 \\ 

\multicolumn{2}{c}{ \ } & \multicolumn{1}{c}{ $\swarrow$ } & 
\multicolumn{1}{c}{ $\searrow$  } & \multicolumn{5}{c}{ \ } & \multicolumn{1}{c}{ $\swarrow$ }
 & \multicolumn{1}{c}{ $\searrow $ } & \multicolumn{2}{c}{ \ }	 \\

\multicolumn{2}{c}{ \ } & \multicolumn{1}{l}{ $\nu $ }  & \multicolumn{2}{r}{ $(1-\nu)$ }
 & \multicolumn{2}{c}{ \ } & \multicolumn{2}{r}{ $\mu $ } & \multicolumn{3}{r}{ $(1-\mu)$ } &
\multicolumn{1}{c}{ \ }	 \\

\multicolumn{1}{r}{ \ } & \multicolumn{1}{c}{ $\swarrow$ } & \multicolumn{2}{r}{ \ } & 
\multicolumn{1}{c}{ $\searrow$  } & \multicolumn{2}{l}{ $ \nearrow $  } & \multicolumn{1}{r}{ $\nwarrow$ } & \multicolumn{1}{c}{ $\swarrow$ } & \multicolumn{2}{r}{ \ } & \multicolumn{1}{c}{ $\searrow $ } &
\multicolumn{1}{l}{ $ \nearrow$  }	 \\

\ldots & $ \ \bullet \ $ & \multicolumn{1}{c}{\ldots} & \ldots & { \ } & \multicolumn{1}{c}{\ldots} &  
\multicolumn{1}{c}{ \ } & 
\ldots & $  \downarrow $  &  \multicolumn{1}{c}{\ldots} &  \ldots & $ \ \bullet \ $ & \ldots \\		

\multicolumn{7}{c}{ \ } &  \multicolumn{1}{c}{\ldots} & \multicolumn{1}{c}{ $ \bullet $ } &  \multicolumn{1}{c}{\ldots} 
 & \multicolumn{3}{c}{ \ }	 \\ 
	
	\end{tabular} 
\end{center}		
An avalanche is represented by a symbol $\stackrel{\boldsymbol{\downarrow}}{\bullet}$. An example of relaxation of the size {\bf three} is presented in the diagram below.
\begin{center} 
		\begin{tabular}{l c|c|c|c|c|c|c|c|c|c|c|c|c|c}
		\multicolumn{9}{c}{ \ } &  \multicolumn{1}{c}{$  \stackrel{\boldsymbol{\downarrow}}{\bullet} $}&  \multicolumn{5}{c}{\ }		\\
		 time $= t $   &  $\quad \quad \cdots$ & $ \bullet $ & $ \bullet $ & $ \ $ & $ \ $ & $ \ $ & $ \bullet $ & $\bullet$ &
			$ {\bullet} $ & $ \ $ & $ \ $ & $ \bullet $ & $ \ $ &$ \cdots \quad \quad $	   \\	
\multicolumn{15}{c}{ \ }		\\
		 time $= t + 1 $  &  $ \quad \quad \cdots$ & $\bullet$ & $\bullet$ & $ \  $ & $ \ $ & $\ $ & $\boldsymbol\downarrow $ &
$\boldsymbol\downarrow $ & $\boldsymbol\downarrow$ &  $ \ $ & $ \ $ & $\bullet$ & $ \ $ & $\cdots \quad \quad$    \\
\multicolumn{7}{c}{ \ } &  \multicolumn{1}{c}{$ {\bullet} $}& \multicolumn{1}{c}{$ {\bullet} $} & 
\multicolumn{1}{c}{$ {\bullet} $} &  \multicolumn{5}{c}{\ }		\\
			\multicolumn{15}{c}{ \ }		\\
		\end{tabular} 
\end{center}		

The name 'domino automaton' comes from the following interpretation. Occupied cells are represented by standing domino blocks and empty cells are represented by extra space between them. If the incoming ball hits an empty space, a domino block is added there with probability $\nu$. If the incoming ball strikes a domino block, with probability $\mu$ it falls down the chosen block and all its adjacent neighbours (on both sides) up to the gap (empty cell). Then thefallen dominoes are removed and the procedure repeats. 

Presented defining rules of the domino automaton can be easily modified leading to various extensions of the system.
We mention some possibilities (like geometry, capacity of cells, many kinds of balls, different triggering rules) in section \ref{sec:concl}. However, the aim of this work is to provide detailed description of the "core" case described above.   

\section{Equations of random domino automaton} \label{sec:eqs}

\subsection{Notation} 
A state of the automaton is defined by altered sequences of occupied and empty cells. 
A sequence of $i$ consecutive occupied cells is called 
the cluster of length $i$ (shortly $i$-cluster); 
a sequence of $i$ consecutive empty cells is called empty cluster of length $i$, where $i=1,2,\ldots,N$. 
The fixed size of the lattice $N$ is assumed to be finite, but big enough to make  
limit for the size of clusters negligible.
Denote the number of $i$-clusters by $n_i$, the number of empty $i$-clusters by $n_i^0$,
the total number of clusters by $n$, and the density (a number of occupied cells divided by the number of all cells $N$) by $\rho$.  
Then, it follows that
\begin{equation} \label{eq:n_rho}
n=\sum_{i\geq 1} n_i = \sum_{i\geq 1} n_i^0 \quad \text{and} \quad \rho = \frac{1}{N} \sum_{i\geq1}  n_i i.
\end{equation}

\begin{figure}[t]
	\centering
	\includegraphics[height=3cm, width=8cm]{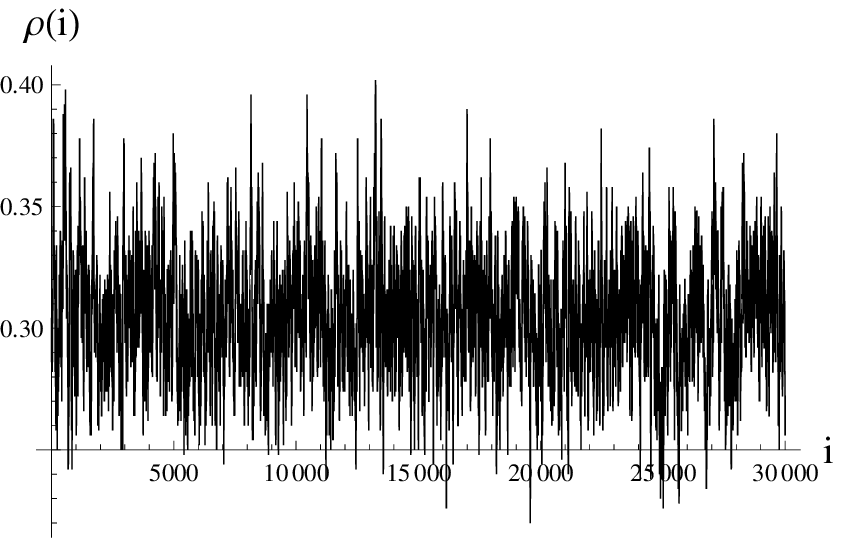} 
	\includegraphics[height=3cm, width=8cm]{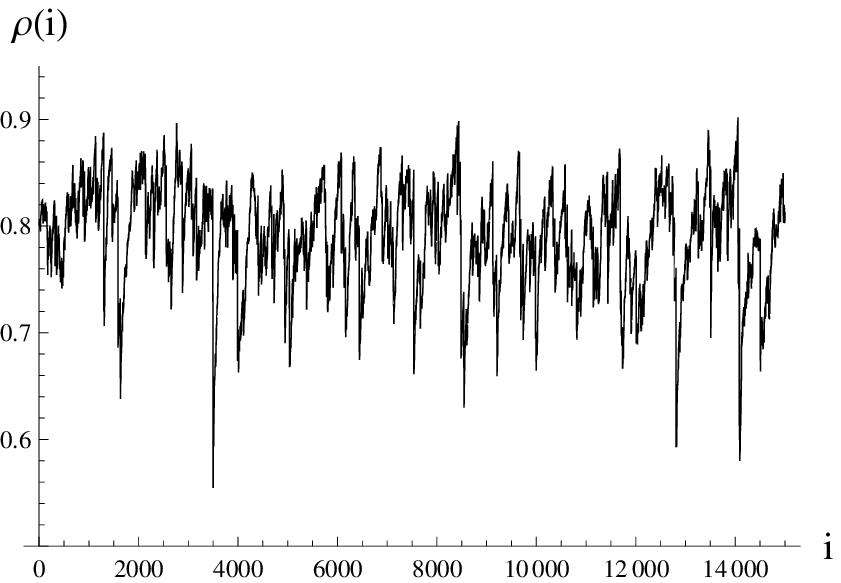} 
	\caption{Examples of simulation results for time series of density $\rho(i)$ of the 1D domino automaton 
in case $\mu/\nu=1$ with lattice size $N=500$ and in case $\mu/\nu=\frac{0.25}{i}$ with lattice size $N=4000$. 
The parameter $i$ numbers avalanches.} \label{fig:Fig1}
\end{figure}

For higher densities $\rho$, the probability of relaxation is also relatively higher, and the density is more likely to decrease.
For smaller densities, triggering of an avalanche is relatively less probable and the density tends to grow up. 
Hence, the variations of the density (see Fig.\ref{fig:Fig1}) are subjected to 'v-shape' potential. 
The behaviour of the automaton is described below under the assumption that 
it is in a quasi-equilibrium state and the variables used in the equations, like density and others, 
are average values and do not depend on time. 
The applicable description of the system as a Markov process is postponed to another paper. 
 
We consider the following settings for rebound parameters. 
The coefficient $\mu=\mu_i$ does not depend on the position of the chosen cell, 
but may depend on the size $i$ of the chosen cluster. The parameter $\nu$ is fixed to be constant.
There are two cases of special interest (discussed in detail in Section \ref{sec:speccases}).
One with $\mu=\beta=\text{const}$ refers to equal probability of provoking an avalanche for each occupied cell, 
while case $\mu=\mu_i=\frac{\delta}{i}$, where ${\delta=\text{const}}$, refers to equal probability of triggering an avalanche for each cluster,  irrelevantly to its length. 
These two respective cases are related to exponential and long-tail distributions of clusters. 
  
\subsection{The balance equation for $\rho$.} 
The stationarity condition requires the flow-in to be equal to flow-out, hence there must be a balance between losses and gains in the number of occupied cells. In a single time step, the number of occupied cells may stay unchanged (when ball is reflected), or may increase by one (an empty cell become occupied), or decrease by $i$ (in a case of an avalanche of size $i$). The expected  value of increase of density is equal to the probability that an empty space becomes occupied, namely $\nu(1-\rho)$.
The probability of relaxation of a size $i$ is $\mu_i \frac{i n_i}{N}$. Since any possible size $i$ can trigger an avalanche, the stationarity gives
\begin{equation} \label{eqn:equil_ni}
\nu(1-\rho) = \frac{1}{N} (\sum_{i\geq1} \mu_i n_i i^2) \quad \text{or} \quad N=
\sum_{i\geq1} n_i i (\frac{\mu_i}{\nu} i+1).
\end{equation}

\subsection{The balance equation for $n$.}
In a single time step, the number of clusters may increase only when a new $1$-cluster is created. 
The chance is proportional to the number of interior cells in empty clusters having the length three and bigger, as depicted in the diagram below.
$$ \cdots | \bullet | \underbrace{ \quad \ | \overbrace{\quad | \cdots \cdots| \quad }^{(i-2) \ = \ \text{interior}} | \ \quad }_{i} | \bullet | \cdots $$
Hence, the probability is 
\begin{equation}
\sim  \sum_{i\geq3} \nu \frac{(i-2)}{i} \frac{n^0_i i}{N} = \nu \left((1-\rho)  - 2 \frac{n}{N} + \frac{n_1^0}{N}\right).
\label{eq:gain1}
\end{equation}

Losses in the total number of clusters may appear in two ways: by joining of two clusters (separated by an empty $1$-cluster) 
and by triggering an avalanche. In the first case, the probability is $\sim \nu \frac{n_1^0}{N}$, in the second it is 
$\sim \sum_{i\geq1} \mu_i \frac{n_i i}{N}$.
Hence, the balance equation for the number of clusters is of the form
\begin{equation}
(1-\rho){N} - 2n  =  \sum_{i\geq1} \frac{\mu_i}{\nu} n_i i.
\label{eq:ballance_n}
\end{equation}

\subsection{Balance equations for $n_i$s.}
We underline the assumption used in the derivation of the equations below: {\it clusters are distributed independently},
by which we mean that the length of the "next" cluster does not depend on the length of the "previous" one. 
In other words, our investigations are done up to the order of clusters (and order of empty clusters).
The presented approach may be regarded as a generalization of percolation, where subsequent {\it cells} are treated as independent. Below, in subsection \ref{ssec:muconst}, we compare both approaches numerically.    

To write down equations for the numbers of clusters of length $i$, i.e., for $n_i$s, we consider 
all possibilities of losses of such clusters as well as creation of them, and next we claim that on the average the gains and losses of respective values compensate each other, as required by the stationarity conditions.

{\bf Losses.} There are two ways to destroy an $i$-cluster: by enlarging and by provoking the avalanche depending on the cell where an incoming particle is thrown. 
$$ \cdots | \quad | \underbrace{ \ \bullet \ | \bullet | \cdots | \ \bullet \ }_{i} | \quad | \cdots $$
{\bf (a)} Enlarging. For any cluster there are two cells adjacent to its ends, so the probability is 
$$ \sim 2 \nu \frac{n_i}{N}.$$
If the single empty cell is between two clusters of the length $i$ it is counted twice - it decrease the number of $i$-clusters by two.   \\
{\bf (b)} Relaxation. In this case it is enough to knock out any of the occupied cell of the cluster,
so the probability is  
$$ \sim \mu_i \frac{i n_i}{N}.$$

{\bf Gains.} There are in general two possibilities to create $i$-cluster: enlarging  $(i-1)$cluster 
and joining two smaller clusters.\\
{\bf (a)} Enlarging. 
Case $i=1$ was already considered and the probability is given by formula \eqref{eq:gain1}.
For $i\geq2$ enlarging a $(i-1)$-cluster to the size $i$ is possible if the adjacent empty cluster is of a size bigger then one. Hence the probability is  
$$ \sim 2 \nu \frac{n_{i-1}}{N} \frac{\sum_{i\geq2} n_i^0}{n^0} 
= 2 \nu \frac{n_{i-1}}{N} \left( 1- \frac{n_1^0}{n} \right). $$
where the multiplier $2$ counts left and right cases. \\
{\bf (b)} Joining two clusters. Two smaller clusters: one of size $k\in \{1,2,\ldots,(i-2)\} $ and the other of size $(i-1)-k$ will be joined if the ball fills an empty cell between them.
$$ \cdots | \quad | \underbrace{ \overbrace{\bullet \ | \cdots | \ \bullet }^k | \quad | 
\overbrace{ \bullet \ | \bullet | \cdots | \ \bullet}^{(i-1-k)}  }_{i} | \quad | \cdots $$
The probability is proportional to the number of empty $1$-clusters between $k$-cluster and $(i-1-k)$-cluster, hence
$$ \sim \nu \frac{n_1^0}{N} \sum_{k=1}^{i-2} \frac{n_k}{n} \cdot \frac{n_{i-1-k}}{n}. $$
The dot in the multiplication above underlines the independence assumption for the order of clusters. 
The last formula introduces also an extra quadratic nonlinearity into the system; so far, nonlinearity 
was present through $n$ (and also $n_1^0$).

Finally the following set of equations for $n_i$s is derived
\begin{eqnarray} 
n_1&=&\frac{1}{\frac{\mu_1}{\nu}+2}\left((1-\rho)N  - 2 n + n_1^0 \right), \label{eq:n1}\\
n_2 &=& \frac{2}{2\frac{\mu_2}{\nu}+2} \left( 1- \frac{n_1^0}{n} \right) n_1 , \label{eq:n2} \\
 n_i &=& \frac{1}{\frac{\mu_i}{\nu} i+2} \times \nonumber \\
 && \times \left( 2 n_{i-1}  \left( 1- \frac{n_1^0}{n} \right) 
+ n_1^0 \sum_{k=1}^{i-2} \frac{n_k n_{i-1-k}}{n^2} \right) \label{eq:ni}  
\end{eqnarray}
for $i\geq 3$, where  $n= \sum_{i\geq1} n_i$ and $\rho = \frac{1}{N}\sum_{i\geq1} i n_i$. 

Variable $n_1^0$ can be cancelled from the above set by
considering the balance for empty clusters and deriving, in analogous way,  
the respective set of equations for variables $n_i^0$ (see Appendix \ref{app:empty_cl_derv} for details)).
The result is  given by equation \eqref{eq:dynn_12^0}, namely
\begin{equation}
n_1^0  = \frac{2 n}{\left(3 + \frac{2}{\nu n} {\sum_{i\geq1} \mu_i n_i i } \right)}.
\label{eq:n10}
\end{equation}
The above, substituted into equations \eqref{eq:n1}-\eqref{eq:ni},
form a closed set for variables  $ \{ n_1, n_2, \ldots \}$. 
Then, equations \eqref{eq:dynn_12^0} and \eqref{eq:dynn_k^0} allow us to find $n^0_k$ for all $k=1,2,\ldots$.

In conclusion, we obtained closed, nonlinear set of equations
for the distribution of the clusters in the cellular automaton. 

\subsection{Moments} \label{ssec:mom}

The balance equations \eqref{eqn:equil_ni} and \eqref{eq:ballance_n} 
(as well as equations for higher weighted moments of $n_i$) can be also obtained from the above set 
\eqref{eq:n1}-\eqref{eq:ni}.

For $n_i$, $i=1,2,\ldots$, we define a moment of order $\gamma$ by
\begin{equation} 
m_\gamma = \frac{1}{N} \sum_{i\geq 1} n_i i^\gamma.
\label{eq:momdef}
\end{equation}
Then, the density $\rho$ is equal to the first moment $m_1$, and the normalized number of clusters $\frac{n}{N}$ is equal to the zero moment $m_0$. 
We define also {\it weighted} moments $\widehat{m}_\gamma$ as follows
\begin{equation} 
\widehat{m}_\gamma = \frac{1}{N} \sum_{i\geq 1} \frac{\mu_i}{\nu} n_i i^\gamma.
\label{eq:wmomdef}
\end{equation}
In this notation, equation \eqref{eq:n10} takes the following form
\begin{equation}
n_1^0  = \frac{2 m_0 N}{3 + 2 \frac{\widehat{m}_1}{m_0} }.
\label{eq:n10m}
\end{equation}
Removing denominators from the set of equations \eqref{eq:n1}-\eqref{eq:ni}, multiplying
both sides by $i^z$ and performing summation with respect to $i$, one obtains
\begin{eqnarray}
\sum_{i\geq1} \frac{\mu_i}{\nu} i^{z+1} n_i + 2\sum_{i\geq1} i^z n_i =  (1-\rho)N -2n +n_1^0 +  \quad \label{eq:momz1} \\
+ 2 (1-\frac{n_1^0}{n}) \sum_{i\geq1} (i+1)^z  n_i +   
 \frac{n_1^0}{n^2} \sum_{i\geq3}\sum_{k=1}^{i-2} i^z n_k n_{i-1-k}. \nonumber
\end{eqnarray}
Finally, after using formula \eqref{eq:n10m}, changing the order of summation (in indices $i$ and $k$), introducing variable $j$ by formula $i=k+1+j$,
expanding binomials and performing sums, one obtains 
\begin{eqnarray}
\widehat{m}_{z+1}&=&1-m_1-2m_0+ 2\sum_{k=0}^{z-1} {z \choose k} m_k +    \quad \label{eq:momfinal} \\
&& + \frac{2}{3m_0+2\widehat{m}_1} \sum_{l,p=1}^{l+p\leq z}  {{z} \choose {l+p}} {{l+p} \choose {l}} m_l m_p. \nonumber
\end{eqnarray}
The above equation for $z=0$ and $z=1$: 
\begin{eqnarray}
\widehat{m}_1 &=& 1-m_1-2m_0, \label{eq:m0s} \\
\widehat{m}_2 &=& 1-m_1,   \label{eq:m1s}
\end{eqnarray}
are the balance of $n$ equation \eqref{eq:ballance_n} and 
the balance of $\rho$ equation \eqref{eqn:equil_ni}, respectively.
Equations of moments for subsequent values of $z$ allow to choose particular forms of rebound parameters
and isolate interesting cases as presented below for formula \eqref{eq:effpardef}.
 

Definition of moments \eqref{eq:momdef} leads to particularly neat 
form of expressions for the average size of a cluster
\begin{equation}
<i> = \frac{\sum_{i\geq1} n_i i}{\sum_{i\geq1} n_i} = \frac{m_1}{m_0},
\label{eq:avrcl}
\end{equation}
and the average size of an avalanche 
\begin{equation}
<w> = \frac{\sum_{i\geq1} \mu_i n_i i^2}{\sum_{i\geq1} \mu_i n_i i} = \frac{\widehat{m}_{2}}{\widehat{m}_{1}}
=\frac{1-m_1}{1-m_1-2m_0}.
\label{eq:avrav}
\end{equation}
 
For rebound parameters in the form
\begin{equation}
\mu_i = \frac{\delta}{i^\sigma}, \quad \quad \nu=const.,
\label{eq:effpardef}
\end{equation} 
where $\delta$, $\sigma$ and $\theta=\frac{\delta}{\nu}$ are constants,
equations \eqref{eq:m0s} and \eqref{eq:m1s} take the form
\begin{eqnarray}
m_1 + \theta m_{2-\sigma} &=& 1,   \label{eq:mombalrho}\\
2m_0+m_1+\theta m_{1-\sigma} &=& 1. \label{eq:mombaln}
\end{eqnarray}
Hence, cases $\sigma=0$ and $\sigma=1$, considered in details below, 
are very suitable for exact analysis. 
There is also an interesting behaviour when $\sigma > 2$, but we consider it in another paper. 

\section{Special cases} \label{sec:speccases}

\subsection{Case $\mu=\beta=const$: equal probability of triggering relaxation for each occupied cell.} \label{ssec:muconst}
To get the value of density from equation \eqref{eqn:equil_ni} in the just considered case, extra relations are required. 
As the first step, consider the percolation approximation \cite{StaufferBook}
\begin{equation} \label{eqn:clind}
n_i= c (1-\rho)^2 \rho^i,
\end{equation}
which means that cells are treated as independent.
From equation \eqref{eq:n_rho} it follows that $c=N$. 
Then equation \eqref{eqn:equil_ni} gives
$\rho={2} / {(3\frac{\beta}{\nu}+ \sqrt{9{(\frac{\beta}{\nu})}^2+4(1-\frac{\beta}{\nu})} )} $.
Any value of density $\rho \in (0,1)$ can be obtained  
for suitable $\frac{\beta}{\nu} \in (0,\infty)$. 
The percolation approximation in the case
$\frac{\beta}{\nu}=1$ gives the value of $\rho=\frac{1}{3}$, which differs by several percent 
from the value $<\rho> \simeq 0.3075$ obtained from numerical simulation of the automaton. 
This divergence appears because the automaton rule states that there is a "coupling" between the adjacent cells, since the relaxations takes out the whole cluster, so treating cells as independent, like in formula \eqref{eqn:clind}, gives 
only very rough  estimation of the density $\rho$. 

In the general case (not percolation approximation), $\mu=\beta=\text{const}$, the balance equation for $n$
gives the following formula for an average size of the cluster 
\begin{equation}
<i> = \frac{N \rho}{n}= \frac{2\rho}{1-\rho(1+\frac{\beta}{\nu})},
\label{eq:avi}
\end{equation} 
and the balance of $\rho$ gives an average size of the avalanche 
\begin{equation}
<w> = \frac{\nu}{\beta} \left(\frac{1-\rho}{\rho}\right).
\label{eq:avw}
\end{equation}

\begin{figure}[t]
	\centering
	\includegraphics[height=4cm, width=8cm]{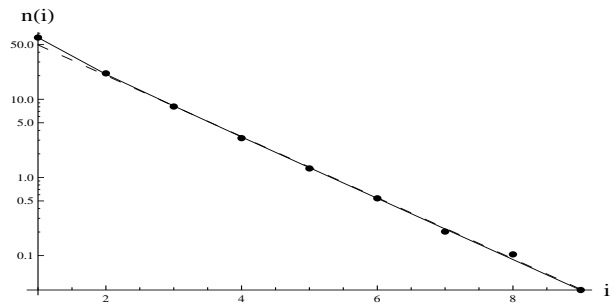} 
	\caption{Values of $n_i$  in case $\mu=\beta=const$ from simulation (dots) compared with values computed from equations
	(solid line) and the approximation $ke^{-\gamma i}$ (dashed line). 
	Lattice size $N=500$ and $\beta/\nu=1$.}
	\label{fig:FigA}
\end{figure}

A comparison of values of the average density $<\rho>$, the average size of cluster $<i>$ and the average size of the avalanche $<w>$ in exemplary case $\mu/\nu=1$ obtained from numerical solution of equations, from percolation approximation results (based on the density balance), and from simulation is presented in Table~\ref{tab:Table1}.
The lattice size is fixed as $N=500$.
 
\begin{table}[t]
\caption{\label{tab:Table1} The average density $<\rho>$, the average size of a cluster $<i>$, and the average size of an avalanche $<w>$ from simulation results, equations of the model and percolation approximation for case $\mu/\nu=1$.}
\begin{ruledtabular}
\begin{tabular}{cccc}
 & Simulation  &  Equations  & Percolation\footnote{Based on the density balance.}  \\
\hline
$<\rho>$ &  0.3075 & 0.308 & 1/3 \\
			$<i>$    &  1.5974 & 1.597 & 3/2 \\
			$<w>$    &  2.2487 & 2.252 & 2  
\end{tabular}
\end{ruledtabular}
\end{table} 
 
Approximate solution for $n_i$s in the case $\mu=\beta=const$ can be obtained
by substituting in equation \eqref{eq:ni} the following formula   
\begin{equation}
n_i=k e^{-\gamma i} \quad \text{for} \quad i=3,4,\ldots
\label{eq:ansatz}
\end{equation}
where $k$ and $\gamma$ are some constants. For $i>>1$ this gives 
\begin{equation}
e^{-\gamma}= \frac{\nu}{\beta}\frac{kn^0_1}{n^2},
\quad \text{or} \quad   
n_i= k \left(\frac{\nu}{\beta} \frac{kn^0_1}{n^2} \right)^i.
\label{eq:n_ilikeper}
\end{equation}
The value of constant $k$ can be found from equation \eqref{eq:n_rho}.
The above result indicates a close relation to the percolation dependence for $n_i$ (see equation \eqref{eqn:clind}). 
Approximation \eqref{eq:n_ilikeper} works well even for $n_2$, as may be seen in Figure \ref{fig:FigA}.

\subsection{Case $\mu=\mu_i=\delta/i$: equal probability of triggering relaxation for each cluster.}
For $\mu_i=\frac{\delta}{i}$, where $ \delta =\text{const}$, the balance for $\rho$ - equation \eqref{eqn:equil_ni} - takes the form 
\begin{equation} \label{eqn:Brho}
\rho=\frac{1}{(\theta+1)}
\end{equation}
where $\theta=\frac{\delta}{\nu}$, which relates the density with the ratio  of coefficients  $\delta$ and $\nu$ only.  
Since $\theta \in (0,\infty)$, any density $ \rho \in (0,1)$ may be realized.
The balance for $n$ - equation \eqref{eq:ballance_n} - is reduced to $(1-\rho){N} = (2+\theta)n.$ 
Together equations \eqref{eqn:equil_ni} and \eqref{eq:ballance_n} give
\begin{equation} \label{eqn:Bn}
n=N\frac{\theta}{(\theta+1)(\theta+2)}.
\end{equation}
The maximal number of clusters is $(1+\sqrt{2})^{-2}N \approx 0.17 N$, as obtained for $\theta=\sqrt{2}$.

\begin{figure}[t]
	\centering
	\includegraphics[height=4cm, width=8cm]{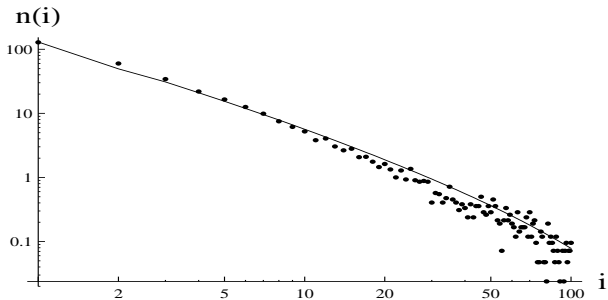} 
	\caption{Exact values of $n_i$  obtained from the model (line) in case  $\mu_i=\delta/i$  compared  
	with exemplary simulation data (dots). 
	Lattice size $N=4000$ and $\delta/\nu={1}/{4}$.}
	\label{fig:FigB}
\end{figure}

\begin{table}[t]
\caption{\label{tab:Table2} The average density $<\rho>$ and the average size of a cluster/avalanche $<i>=<w>$ from simulation results and equations of the model for the case $\mu_i=\frac{\delta}{i}$, for $\theta=1/4$.}
\begin{ruledtabular}
\begin{tabular}{ccc}
 & Simulation  &  Equations   \\
\hline
$<\rho>$ &  0.7926 & 0.8  \\
$<i>=<w>$    &  8.967 & 9
\end{tabular}
\end{ruledtabular}
\end{table} 

The average size of a cluster is the same as the average size of an avalanche, because  
any cluster, regardless of its size, has the same chance to become an avalanche. 
Hence, both balance equations lead to the same result
\begin{equation*}
<i> = <w> = 1 + \frac{2}{\theta}.
\label{eq:avicl}
\end{equation*}
Table \ref{tab:Table2} contains values of average density and average cluster/avalanche size 
obtained from equations and simulation in an exemplary case where $\theta={0.25}$. 
The size of lattice is set to $N=4000$ in order to avoid restrictions for size of avalanches,
which are relatively bigger comparing to the previous case.

A simple form of $n_1^0  = {2 n}/{\left(3 + 2\theta \right)}$ 
together with balance equations \eqref{eqn:Brho} and \eqref{eqn:Bn} alows to  
reduce the set of equations \eqref{eq:n1}-\eqref{eq:ni} to the solvable recurrence.
The formulas for $n_i$s are rational functions of  $\theta$ and $N$ only.
The explicit form of solutions and relation to the Motzkin numbers is presented in \cite{BiaMotzkin}.

Figure \ref{fig:FigB} presents exact values of $n_i$s compared in log-log scale
with the simulation result in the exemplary case described above. 
The slope of the distribution of $n_i$ approximates inverse-power distribution in the middle part of a range of sizes $i$;
for bigger values of $i$, values of $n_i$ decrease quicker. 
This property agrees with convergence of moments of $n_i$ we use above.  
For value of $\theta$ closer to $0$ the straight part of the plot extends for bigger values of $i$.   
In the limit case $\theta \stackrel{\theta N = const}{\longrightarrow} 0$, 
the distribution is given by the following power law \cite{BiaMotzkin}
\begin{equation}
n_{i+1} \sim \frac{1}{i^{\frac{3}{2}}}.
\label{eq:powerlaw}
\end{equation}

\section{Conclusions.} \label{sec:concl}

We proposed and studied properties of simple stochastic cellular automaton with avalanches
inspired by an extremely simplified model of earthquakes. The main goal was to construct a simple model 
with "transparent" mathematical structure, which would enable obtaining exact results yet covering a wide range of behaviours.
The rule of the automaton allows to derive from the first principles (using elementary combinatorics) 
the set of equations describing the average values of the model parameters. 
These equations leads to equations for moments and to neat formulas for average cluster and avalanche sizes in terms of the zero and the first moments. 
  
After analysis of moment equations, we considered in detail two cases which differ by the form of parameter $\mu$ responsible for triggering of avalanches.
We obtained exponential type and inverse-power type distributions of clusters within uniform framework. 
(The distribution of avalanches $w_i$ is easily obtained from the formula 
$w_i=\frac{\mu_i n_i i}{\sum_{i\geq1}\mu_i n_i i}\sim \mu_i n_i i$.)
The quasi-equilibrium assumption seems to be justified by simulation results.

The exponential case $\mu=\beta=const$ is not fully analytically solvable; nevertheless, the closed set of equations
leads to several exact relations. 
Moreover, an approximate formula for $n_i$ was derived, and intimate relation to percolation was pointed out. Hence, the proposed approach can be regarded as an extension of percolation approximation results, and can provide basis for exact treatment of other models.  The automaton rule clearly states that there is a "coupling" between the adjacent cells, since the relaxations takes out the whole cluster, so treating cluster -- not cells -- as independent is a substantial improvement in the presented case.
 
The case $\mu_i=\frac{\delta}{i}$ is fully solvable, and all variables are expressed as rational functions of ratio $\frac{\delta}{\nu}$ of rebound parameters only. The shape of the distribution of clusters $n_i$ approximate 
an inverse-power distribution in the various ranges of variable $i$, depending on chosen parameters. In the limit case, distribution tends to $\sim n^{-\frac{3}{2}}$.  The critical density is equal to $1$.
The limit case of the presented automaton reduces the set of equations to the recurrence, which 
leads to known integer sequence - the Motzkin numbers (see \cite{BiaMotzkin}).
This result establishes a new, remarkable link between the combinatorial object and the stochastic cellular automaton. 

The first application of RDA - related to Ito equation - 
is already studied in our parallel papers \cite{CzBiaTL, CzBiaEF}. 
The model serves as a fully controlled stochastic "phenomenon" and an applicability of the 
reconstruction of the Ito equation from generated time series was tested.
Due to its simplicity, it allows to derive exactly the suitable Ito equation, and analytical results were compared with 
histogram method. 
The obtained results are part of broader studies of the privilege concept and its 
role for appearance of inverse-power distributions \cite{Cz03, CzRoz}.

The model posses also some properties which may be used in seeking of applications to natural phenomena and 
checking their adequacy.  
For example, in the context of earthquakes, the discrete nature of the model makes possible an adjustment of positions and sizes of cells to the geological structure of a fault. Bending and shrinking of the 1-D grid
have no influence on equations; the only important feature is an order of cells. In this context, relations describing average size of cluster $<i>$ and avalanche $<w>$ may be interpreted as relations giving their dependence on the density of energy on the wedge. 

The model establishes one-to-one correspondence between  distribution of avalanches (i.e. observable quantity) and respective parameter $\frac{\mu_i}{\nu}$ responsible for triggering avalanches, as it was illustrated in two cases considered above. 
We emphasize that the relation works in both directions. 
Fixing rebound parameters leads to various distributions of avalanches, but also, for any distribution $n_i$, one can 
easily find unique values of rebound parameters $\frac{\mu_i}{\nu}$ by simple use of the set of equations \eqref{eq:n1}-\eqref{eq:ni}.
Thus, within the proposed model, one can infer about triggering properties from "observed" statistics of avalanches. 
For an earthquake interpretation, it would give an insight into properties of microscopic mechanism of releasing energy on a fault on the basis of already collected data.

Last but not least, we propose few generalizations possibly leading towards more realistic extensions of 
the model.
Each element of the presented automaton - including the incidence rule  - can be subjected to various modifications. 
To be more specific, one can consider different geometry of the array (for example, a tree shaped like Bethe lattice
or any in bigger dimension), different capacities of cells and different kinds of balls distributed to the system. Also there are many kinds of dependence of energy release threshold on other parameters (on space position, on states of cells in a neighbourhood etc.). We leave these topics for further investigations.

\appendix

\section{Derivation of balance equations for empty clusters.} \label{app:empty_cl_derv}
In analogy to the set of equations \eqref{eq:n1}-\eqref{eq:ni} for clusters, one can deduce the following set of equations for the empty clusters. 

{\bf Losses.} There are two following possibilities.\\
{\bf (a)} Occupation of any cell belonging to the empty $k$-cluster. The probability is just $$\sim \nu \frac{k n_k^0}{N}.$$ 
{\bf (b)} Provoking an avalanche. An empty cluster of the length $k$ is lost when one of the two adjacent clusters is knocked out and forms an avalanche.  
$$ \cdots | \quad | \overbrace{  \bullet \ | \bullet | \ \bullet }^{i} | 
\overbrace{\quad  | \quad | \quad | \quad | \quad | \quad}^{k}  | \bullet | \cdots $$
The probability is
$$ \sim 2 \frac{n_k^0}{n^0} \sum_{i\geq1} \mu_i \frac{i n_i}{N}.$$ 
It comes from the probability of knocking out of $i$-cluster times the probability that on the end of the cluster is empty $k$-cluster, times two ends, summed for all possible values of $i$.  

{\bf Gains.} Again, there are two possibilities.\\
{\bf (a)} Shortening an empty cluster. An empty cluster of a length $k$ can be obtained from a bigger one of the length $i \geq k+1$
when an incoming ball hits its $(k+1)$th empty cell 
$$ \cdots | \bullet | \overbrace{\underbrace{\quad | \quad | \quad }_{k}|  \times | \quad | \quad | \quad  | \quad | \quad}^{i} | \bullet | \cdots $$
or symmetrically, its $(i-k)$th cell. Hence the relevant gain term is of the form 
$$ \sim 2 \nu \sum_{i\geq k+1 } \frac{n_i^0}{N}.$$
{\bf (b)} Provoking an avalanche. If an incoming ball knock out the cluster of the length $j \in \{ 1,2,\ldots,k-2 \}$ surrounded by two empty clusters of the respective lengths $l \in \{1,2,\ldots,k-1-j\}$ and $(k-j-l)$, 
$$ \cdots | \bullet | \underbrace{ \overbrace{\quad | \quad }^l | \overbrace{  \bullet \ | \bullet | \ \bullet }^{j} | 
\overbrace{\quad  | \quad | \quad | \quad}^{k-l-j}  }_{k} | \bullet | \cdots $$
then the empty cluster of the length $k$ is created. Therefore, for $k \geq 3$ this leads to the following term
$$ \sim \sum_{j=1}^{k-2} \sum_{l=1}^{k-1-j} \mu_j \frac{j n_j}{N} \frac{n^0_l}{n^0}\frac{n^0_{k-j-l}}{n^0}.$$ 
This is the probability of knocking out of the $j$-cluster times probability the $l$-empty cluster is on the left times 
probability the empty $({k-l-j})$-cluster is on the right, for all possible values of $l$.

{\bf Balance equations.} The set of equations for the average values in the quasi equilibrium is of the form 
\begin{eqnarray} 
n_k^0 \left( k + \frac{2}{n}\sum_{i\geq1} \frac{\mu_i}{\nu} i n_i \right)  = 2 \sum_{i\geq k+1} n_i^0,  &&    k=1,2, 
\label{eq:dynn_12^0} \\
n_k^0 \left( k +  \frac{2}{n}\sum_{i\geq1} \frac{\mu_i}{\nu} i n_i \right) = 2 \sum_{i\geq k+1} n_i^0 && +   \nonumber \\
 + \sum_{j=1}^{k-2}  \sum_{l=1}^{k-1-j}  \frac{\mu_j}{\nu} \frac{j n_j}{N} \frac{n^0_l}{n^0}\frac{n^0_{k-j-l}}{n^0},   
&&  k\geq 3. \label{eq:dynn_k^0}
\end{eqnarray}  
Equation \eqref{eq:dynn_12^0} may be written as  
\begin{equation*}
n_k^0 = \frac{ (k+2) + \frac{2}{n}\sum_{i\geq1} \frac{\mu_i}{\nu} i n_i } 
 {2 \left( n - \sum_{i=1}^{k-1} n_i^0  \right)} \quad \quad \text{for} \quad \quad  k=1,2,
\end{equation*} 
and equation \eqref{eq:dynn_k^0}, valid for $ k\geq 3 $ is
\begin{equation*}
n_k^0  = \frac{ (k+2) + \frac{2}{n}\sum_{i\geq1} \frac{\mu_i}{\nu} i n_i } 
 {2 ( n - \sum_{i=1}^{k-1} n_i^0 ) + \sum_{j=1}^{k-2}  \sum_{l=1}^{k-1-j} \frac{\mu_j}{\nu} \frac{j n_j}{N} \frac{n^0_l}{n^0}\frac{n^0_{k-j-l}}{n^0} }.
\end{equation*}
Thus, on right hand sides there are terms $n^0_j$ with $j < k$ only.

{\bf Balance of $n$ out of empty clusters.} 
The set of equation for empty clusters can be treated by methods we use in
Subsection \ref{ssec:mom} in order to derive relations for respective moments.

Below we consider zero moment for $n_i^0$, which leads also to equation \eqref{eq:ballance_n}. 
This is implied by the fact that the number of empty clusters is equal to 
the number of clusters (see equation \eqref{eq:n_rho}).
To perform sum of \eqref{eq:dynn_12^0} and \eqref{eq:dynn_k^0} for all values of $k$ notice the following identities:
$$ \sum_{k\geq1}  n_k^0 \left( k + \frac{2}{n}\sum_{i\geq1} \frac{\mu_i}{\nu} i n_i  \right) = 
(1-\rho) N + 2\sum_{i\geq1} \frac{\mu_i}{\nu} i n_i, $$
$$ \sum_{k\geq1} \sum_{i\geq k+1} n_i^0 = \sum_{k\geq1} (i-1) n_i^0 = (1-\rho)N - n,$$ 
and 
$$ \sum_{k\geq 3} \sum_{j=1}^{k-2} \sum_{l=1}^{k-1-j} \frac{\mu_j}{\nu} \frac{j n_j}{N} \frac{n^0_l}{n^0}\frac{n^0_{k-j-l}}{n^0} = \frac{1}{N}\sum_{j\geq1} \frac{\mu_j}{\nu}{j n_j}.$$
In the last  identity we assumed that it is allowed to change the order of summation with respect to indices $j$ and $k$.
Thus, a sum of \eqref{eq:dynn_12^0} and \eqref{eq:dynn_k^0} for all $k$ leads directly to 
$$ (1-\rho){N} - 2n  =  \sum_{i\geq1} \frac{\mu_i}{\nu} n_i i.$$ 
It is the balance equation \eqref{eq:ballance_n} already obtained.

\section*{Acknowledgement}
This work was partially supported by the project INTAS 05-1000008-7889.


\begin{thebibliography}{26}
\expandafter\ifx\csname natexlab\endcsname\relax\def\natexlab#1{#1}\fi
\expandafter\ifx\csname bibnamefont\endcsname\relax
  \def\bibnamefont#1{#1}\fi
\expandafter\ifx\csname bibfnamefont\endcsname\relax
  \def\bibfnamefont#1{#1}\fi
\expandafter\ifx\csname citenamefont\endcsname\relax
  \def\citenamefont#1{#1}\fi
\expandafter\ifx\csname url\endcsname\relax
  \def\url#1{\texttt{#1}}\fi
\expandafter\ifx\csname urlprefix\endcsname\relax\def\urlprefix{URL }\fi
\providecommand{\bibinfo}[2]{#2}
\providecommand{\eprint}[2][]{\url{#2}}

\bibitem[{\citenamefont{Tang and Bak}(1988)}]{TangBak-mf}
\bibinfo{author}{\bibfnamefont{C.}~\bibnamefont{Tang}} \bibnamefont{and}
  \bibinfo{author}{\bibfnamefont{P.}~\bibnamefont{Bak}}, \bibinfo{journal}{J.
  Stat. Phys.} \textbf{\bibinfo{volume}{51 No 5/6}}, \bibinfo{pages}{797}
  (\bibinfo{year}{1988}).

\bibitem[{\citenamefont{Dhar and Ramaswamy}(1989)}]{DharExact}
\bibinfo{author}{\bibfnamefont{D.}~\bibnamefont{Dhar}} \bibnamefont{and}
  \bibinfo{author}{\bibfnamefont{R.}~\bibnamefont{Ramaswamy}},
  \bibinfo{journal}{Phys. Rev. Lett.} \textbf{\bibinfo{volume}{63}},
  \bibinfo{pages}{1659} (\bibinfo{year}{1989}).

\bibitem[{\citenamefont{Vespignani and Zapperi}(1998)}]{Vespignani}
\bibinfo{author}{\bibfnamefont{A.}~\bibnamefont{Vespignani}} \bibnamefont{and}
  \bibinfo{author}{\bibfnamefont{S.}~\bibnamefont{Zapperi}},
  \bibinfo{journal}{Phys. Rev. E} \textbf{\bibinfo{volume}{57}},
  \bibinfo{pages}{6345} (\bibinfo{year}{1998}).

\bibitem[{\citenamefont{Alcaraz and Rittenberg}(2008)}]{Alcaraz-DirAbAlg}
\bibinfo{author}{\bibfnamefont{F.C.}~\bibnamefont{Alcaraz}} \bibnamefont{and}
  \bibinfo{author}{\bibfnamefont{V.}~\bibnamefont{Rittenberg}},
  \bibinfo{journal}{Phys. Rev. E} \textbf{\bibinfo{volume}{78}},
  \bibinfo{pages}{041126} (\bibinfo{year}{2008}).

\bibitem[{\citenamefont{Dhar}(1999)}]{DharSanpile}
\bibinfo{author}{\bibfnamefont{D.}~\bibnamefont{Dhar}},
  \bibinfo{journal}{Physica A} \textbf{\bibinfo{volume}{263}},
  \bibinfo{pages}{4} (\bibinfo{year}{1999}).

\bibitem[{\citenamefont{Gaveau and Schulman}(1991)}]{Gaveau-mfsoc}
\bibinfo{author}{\bibfnamefont{B.}~\bibnamefont{Gaveau}} \bibnamefont{and}
  \bibinfo{author}{\bibfnamefont{L.}~\bibnamefont{Schulman}},
  \bibinfo{journal}{J. Phys. A: Math. Gen.} \textbf{\bibinfo{volume}{24}},
  \bibinfo{pages}{L475} (\bibinfo{year}{1991}).

\bibitem[{\citenamefont{Lee et~al.}(1991)\citenamefont{Lee, Liang, and
  Tzeng}}]{detSandpile}
\bibinfo{author}{\bibfnamefont{S.C.} \bibnamefont{Lee}},
  \bibinfo{author}{\bibfnamefont{N.Y.}~\bibnamefont{Liang}}, \bibnamefont{and}
  \bibinfo{author}{\bibfnamefont{W.J.} \bibnamefont{Tzeng}},
  \bibinfo{journal}{Phys. Rev. Lett.} \textbf{\bibinfo{volume}{67}},
  \bibinfo{pages}{1479} (\bibinfo{year}{1991}).

\bibitem[{\citenamefont{Takahashi and Satsuma}(1990)}]{TS-aut}
\bibinfo{author}{\bibfnamefont{D.}~\bibnamefont{Takahashi}} \bibnamefont{and}
  \bibinfo{author}{\bibfnamefont{J.}~\bibnamefont{Satsuma}},
  \bibinfo{journal}{J. Phys. Soc. Jpn.} \textbf{\bibinfo{volume}{59 No 10}},
  \bibinfo{pages}{3514} (\bibinfo{year}{1990}).

\bibitem[{\citenamefont{Tokihiro et~al.}(1996)\citenamefont{Tokihiro,
  Takahashi, Matsukidaira, and Satsuma}}]{TTMS}
\bibinfo{author}{\bibfnamefont{T.}~\bibnamefont{Tokihiro}},
  \bibinfo{author}{\bibfnamefont{D.}~\bibnamefont{Takahashi}},
  \bibinfo{author}{\bibfnamefont{J.}~\bibnamefont{Matsukidaira}},
  \bibnamefont{and} \bibinfo{author}{\bibfnamefont{J.}~\bibnamefont{Satsuma}},
  \bibinfo{journal}{Phys. Rev. Lett.} \textbf{\bibinfo{volume}{76}},
  \bibinfo{pages}{3247} (\bibinfo{year}{1996}).

\bibitem[{\citenamefont{Tokihiro}(2004)}]{TokiLNP}
\bibinfo{author}{\bibfnamefont{T.}~\bibnamefont{Tokihiro}}, in
  \emph{\bibinfo{booktitle}{Discrete {I}ntegrable {S}ystems}}, edited by
  \bibinfo{editor}{\bibfnamefont{B.}~\bibnamefont{Grammaticos}},
  \bibinfo{editor}{\bibfnamefont{T.}~\bibnamefont{Kosmann-Schwarzbach}},
  \bibnamefont{and}
  \bibinfo{editor}{\bibfnamefont{T.}~\bibnamefont{Tamizhmani}}
  (\bibinfo{publisher}{Springer}, \bibinfo{year}{2004}), vol.
  \bibinfo{volume}{644} of \emph{\bibinfo{series}{Lecture Notes in Physics}},
  pp. \bibinfo{pages}{383--424}.

\bibitem[{\citenamefont{Bia{\l}ecki and Doliwa}(2005)}]{BD-hyp}
\bibinfo{author}{\bibfnamefont{M.}~\bibnamefont{Bia{\l}ecki}} \bibnamefont{and}
  \bibinfo{author}{\bibfnamefont{A.}~\bibnamefont{Doliwa}},
  \bibinfo{journal}{Commun. Math. Phys.} \textbf{\bibinfo{volume}{253}},
  \bibinfo{pages}{157} (\bibinfo{year}{2005}).

\bibitem[{\citenamefont{Bialecki}(2009)}]{BiaRIMS}
\bibinfo{author}{\bibfnamefont{M.}~\bibnamefont{Bialecki}},
  \bibinfo{journal}{RIMS Kokyuroku} \textbf{\bibinfo{volume}{1650}},
  \bibinfo{pages}{154} (\bibinfo{year}{2009}).

\bibitem[{\citenamefont{Doliwa et~al.}(2003)\citenamefont{Doliwa, Bia{\l}ecki,
  and Klimczewski}}]{DBK}
\bibinfo{author}{\bibfnamefont{A.}~\bibnamefont{Doliwa}},
  \bibinfo{author}{\bibfnamefont{M.}~\bibnamefont{Bia{\l}ecki}},
  \bibnamefont{and}
  \bibinfo{author}{\bibfnamefont{P.}~\bibnamefont{Klimczewski}},
  \bibinfo{journal}{J. Phys. A: Math. Gen.} \textbf{\bibinfo{volume}{36}},
  \bibinfo{pages}{4827} (\bibinfo{year}{2003}).

\bibitem[{\citenamefont{Bia{\l}ecki}(2011)}]{BiaMotzkin}
\bibinfo{author}{\bibfnamefont{M.}~\bibnamefont{Bia{\l}ecki}}
  (\bibinfo{year}{2011}), \bibinfo{note}{{\tt{arXiv:1102.0437 [math-ph]}}}.

\bibitem[{\citenamefont{Czechowski and
  Bia{\l}ecki}(2010{\natexlab{a}})}]{CzBiaTL}
\bibinfo{author}{\bibfnamefont{Z.}~\bibnamefont{Czechowski}} \bibnamefont{and}
  \bibinfo{author}{\bibfnamefont{M.}~\bibnamefont{Bia{\l}ecki}}
  (\bibinfo{year}{2010}{\natexlab{a}}), \bibinfo{note}{{\tt{arXiv:1012.5902
  [nlin.CG]}}}.

\bibitem[{\citenamefont{Czechowski and
  Bia{\l}ecki}(2010{\natexlab{b}})}]{CzBiaEF}
\bibinfo{author}{\bibfnamefont{Z.}~\bibnamefont{Czechowski}} \bibnamefont{and}
  \bibinfo{author}{\bibfnamefont{M.}~\bibnamefont{Bia{\l}ecki}}
  (\bibinfo{year}{2010}{\natexlab{b}}), \bibinfo{note}{{\tt{arXiv:1101.0098
  [nlin.CG]}}}.

\bibitem[{\citenamefont{Sornette}(2006)}]{SornetteCritical}
\bibinfo{author}{\bibfnamefont{D.}~\bibnamefont{Sornette}},
  \emph{\bibinfo{title}{Critical {P}henomena in {N}atural {S}ciences, 2nd
  edition}} (\bibinfo{publisher}{Springer}, \bibinfo{year}{2006}).

\bibitem[{\citenamefont{Bhattacharyya and Chkrabarti}(2006)}]{LNP705}
\bibinfo{author}{\bibfnamefont{P.}~\bibnamefont{Bhattacharyya}}
  \bibnamefont{and}
  \bibinfo{author}{\bibfnamefont{B.}~\bibnamefont{Chkrabarti}},
  \emph{\bibinfo{title}{Modelling {C}ritical and {C}atastrophic {P}henomena in
  {G}eoscience}}, vol. \bibinfo{volume}{705} of \emph{\bibinfo{series}{Lecture
  Notes in Physics}} (\bibinfo{publisher}{Springer}, \bibinfo{year}{2006}).

\bibitem[{\citenamefont{Tejedor et~al.}(2008)\citenamefont{Tejedor, Ambroj,
  Gomez, and Pacheco}}]{Pach08}
\bibinfo{author}{\bibfnamefont{A.}~\bibnamefont{Tejedor}},
  \bibinfo{author}{\bibfnamefont{S.}~\bibnamefont{Ambroj}},
  \bibinfo{author}{\bibfnamefont{J.~B.} \bibnamefont{Gomez}}, \bibnamefont{and}
  \bibinfo{author}{\bibfnamefont{A.~F.} \bibnamefont{Pacheco}},
  \bibinfo{journal}{J. Phys. A: Math. Theor} \textbf{\bibinfo{volume}{41}},
  \bibinfo{pages}{375102 (16pp)} (\bibinfo{year}{2008}).

\bibitem[{\citenamefont{Vazquez-Prada et~al.}(2002)\citenamefont{Vazquez-Prada,
  Gonzalez, Gomez, and Pacheco}}]{PachMin}
\bibinfo{author}{\bibfnamefont{M.}~\bibnamefont{Vazquez-Prada}},
  \bibinfo{author}{\bibfnamefont{A.}~\bibnamefont{Gonzalez}},
  \bibinfo{author}{\bibfnamefont{J.~B.} \bibnamefont{Gomez}}, \bibnamefont{and}
  \bibinfo{author}{\bibfnamefont{A.~F.} \bibnamefont{Pacheco}},
  \bibinfo{journal}{Nonlinear Processes in Geophysics}
  \textbf{\bibinfo{volume}{9}}, \bibinfo{pages}{513} (\bibinfo{year}{2002}).

\bibitem[{\citenamefont{Gonzalez et~al.}(2006)\citenamefont{Gonzalez,
  Vazquez-Prada, Gomez, and Pacheco}}]{PachTect}
\bibinfo{author}{\bibfnamefont{A.}~\bibnamefont{Gonzalez}},
  \bibinfo{author}{\bibfnamefont{M.}~\bibnamefont{Vazquez-Prada}},
  \bibinfo{author}{\bibfnamefont{J.~B.} \bibnamefont{Gomez}}, \bibnamefont{and}
  \bibinfo{author}{\bibfnamefont{A.~F.} \bibnamefont{Pacheco}},
  \bibinfo{journal}{Tectonophysics} \textbf{\bibinfo{volume}{424}},
  \bibinfo{pages}{319} (\bibinfo{year}{2006}).

\bibitem[{\citenamefont{Sloane and Plouffe}(1995)}]{SloaneEnc}
\bibinfo{author}{\bibfnamefont{N.}~\bibnamefont{Sloane}} \bibnamefont{and}
  \bibinfo{author}{\bibfnamefont{S.}~\bibnamefont{Plouffe}},
  \emph{\bibinfo{title}{The {E}ncyclopedia of {I}nteger {S}equences}}
  (\bibinfo{publisher}{Academic Press}, \bibinfo{year}{1995}).

\bibitem[{\citenamefont{Inui and Katori}(1996)}]{Inui-Catalan}
\bibinfo{author}{\bibfnamefont{N.}~\bibnamefont{Inui}} \bibnamefont{and}
  \bibinfo{author}{\bibfnamefont{M.}~\bibnamefont{Katori}},
  \bibinfo{journal}{J. Phys. A: Math. Gen.} \textbf{\bibinfo{volume}{29}},
  \bibinfo{pages}{4347} (\bibinfo{year}{1996}).

\bibitem[{\citenamefont{Stauffer and Aharony}(1992)}]{StaufferBook}
\bibinfo{author}{\bibfnamefont{D.}~\bibnamefont{Stauffer}} \bibnamefont{and}
  \bibinfo{author}{\bibfnamefont{A.}~\bibnamefont{Aharony}},
  \emph{\bibinfo{title}{Introduction to {P}ercolation {T}heory}}
  (\bibinfo{publisher}{Taylor {\&} Francis}, \bibinfo{year}{1992}).

\bibitem[{\citenamefont{Czechowski}(2003)}]{Cz03}
\bibinfo{author}{\bibfnamefont{Z.}~\bibnamefont{Czechowski}},
  \bibinfo{journal}{Geophys. J. Int.} \textbf{\bibinfo{volume}{154}},
  \bibinfo{pages}{754} (\bibinfo{year}{2003}).

\bibitem[{\citenamefont{Czechowski and Rozmarynowska}(2008)}]{CzRoz}
\bibinfo{author}{\bibfnamefont{Z.}~\bibnamefont{Czechowski}} \bibnamefont{and}
  \bibinfo{author}{\bibfnamefont{A.}~\bibnamefont{Rozmarynowska}},
  \bibinfo{journal}{Physica A} \textbf{\bibinfo{volume}{387}},
  \bibinfo{pages}{5403} (\bibinfo{year}{2008}).

\end{thebibliography}

\end{document}